# Hardening of Artificial Neural Networks for Use in Safety-Critical Applications – A Mapping Study

Rasmus Adler[1], Mohammed Naveed Akram[1], Pascal Bauer[1], Patrik Feth[3],
Pascal Gerber[1], Andreas Jedlitschka[1], Lisa Jöckel[1], Michael Kläs[1], Daniel Schneider[1]

*rasmus.adler@iese.fraunhofer.de, naveed.akram@iese.fraunhofer.de,
pascalbauer55@googlemail.com, patrik.feth@sick.de, p_gerber14@cs.uni-kl.de,
andreas.jedlitschka@iese.fraunhofer.de, lisa.joeckel@iese.fraunhofer.de,
michael.klaes@iese.fraunhofer.de, daniel.schneider@iese.fraunhofer.de*

[1]Fraunhofer Institute for Experimental
Software Engineering IESE
Fraunhofer-Platz 1, 67663 Kaiserslautern,
Germany

[3]SICK AG
Erwin-Sick-Straße 1,
79183 Waldkirch,
Germany

**ABSTRACT**

**Context:** Across different domains, Artificial Neural Networks (ANNs) are used more and more in safety-critical applications in which erroneous outputs of such ANN can have catastrophic consequences. However, the development of such neural networks is still immature and good engineering practices are missing. With that, ANNs are in the same position as software was several decades ago. Today, standards for functional safety, such as ISO 26262 in the automotive domain, require the application of a collection of proven engineering principles and methods in the creation of software to increase its quality and reduce failure rates to an acceptable level.

**Objective:** In the future, such a set of proven engineering methods needs to be established for the development of Artificial Neural Networks to allow their use in safety-critical applications.

**Method:** This work takes a step in this direction by conducting a mapping study to extract challenges faced in the development of ANNs for safety-critical applications and to identify methods that have been used for the hardening of ANNs in such settings.

**Results:** We extracted ten different challenges found to be repeatedly reported in the literature regarding the use of ANNs in critical contexts. All of these challenges are addressed by engineering methods, of which we identified 54 in our study that can be used for the hardening of networks.

**Conclusions:** Various methods have been proposed to overcome the specific challenges of using ANNs in safety-critical applications. On the path towards defining best practices, we envision that future software engineering will need to focus on further investigating these methods and increasing the maturity and understanding of existing approaches, with the goal to develop clear guidance for proper engineering of high-quality ANNs.

# 1 INTRODUCTION

Different domains, such as robotics, controls, medicine, defense, and automotive are starting to include ANNs in safety-critical applications. Safety-critical applications are those that can cause harm to the environment in which they are used. The reasons for the use of ANNs in such applications mainly include the significant progress made in the field of Machine Learning in the recent past, mostly related to learning and generalization capabilities (1) (2). Progress in computing power and new efficient learning algorithms, including new processing elements, allows for deeper networks with more layers. These networks can process complex data structures such as images and voice as input instead of relying on manual feature extraction. In this way, ANNs are able to fulfill functional requirements that are yet difficult to tackle with traditionally engineered algorithms. Prominent examples are object detection in images, speech recognition and translation, or competing with humans in board and video games.

The use of Artificial Neural Networks in safety-critical applications requires sufficient confidence that these networks will not (unforeseeably) fail. In engineering, the creation of this confidence is generally referred to as "assurance". There are two fundamentally different and potentially complimentary approaches to provide assurances for systems utilizing ANNs: (1) On the one hand, a redundant "supervisory" channel can be established by traditional engineering means. This means that the ANN is removed from the safety-critical path. Here the challenge lies in designing the supervisor in a way that minimizes the negative impact on the overall system performance. (2) On the other hand, the ANN itself might be engineered in such a way that it provides sufficient assurance that it can be part of a safety-critical functionality. Corresponding methods and techniques might generally be applied at any typical engineering stage, from the selection of training data to analysis and validation of the final ANN product, to harden the network and make it fit for its safety-critical purpose.

This review focuses on work addressing the second approach, meaning work that aims at hardening Artificial Neural Networks for use in safety-critical applications.

The data-driven development of an ANN differs significantly from the traditional algorithm-centered approach for the development of software. Additionally, a trained neural network is a lot different from regular source code. It is therefore not easy to transfer the engineering practices from the development of software, such as programming language restrictions or the use of inspection techniques, to the development of ANNs. There is currently no established or generally agreed set of engineering best practices that can guide the development of ANNs and ensure a high level of quality. To be able to continue to benefit from the advantages provided by ANNs, also in safety-critical applications, it is, however, necessary to identify a catalog of best practices.

For the development of safety-critical software, existing safety standards such as ISO 26262 demand the application of known best practices in order to increase the confidence that the failure rate of the developed source code will reach a level that corresponds to the criticality of the functionality realized by that code. Only with that confidence can the software be used. We need to achieve the same for ANNs: building a relationship between the application of engineering methods and the confidence in the sufficient reduction of (unforeseeable) failures. Given such a relationship, we can then stipulate requirements on the development process of an ANN in the form of these engineering methods.

The contribution of this article is a further step towards creating this catalog. To achieve this contribution, we conducted a systematic mapping study. The presented study is based on the methodology described by Petersen et al. (3). Our aim was to assess from the identified literature the particular challenges of ANNs and their development that hinder their use in safety-critical applications. We further extracted the engineering methods that exist to overcome those challenges. In Section 2, we present related work in the field of safety-critical ANNs. In Section 5, we introduce the process of the mapping study we conducted, before providing the results in Section 4 and reflecting on them in Section 5. Section 6 concludes the paper with an outlook on the usage of the results presented here.

# 2 RELATED WORK

To the best of our knowledge, to date no systematic mapping study has been conducted with the intention to collect challenges and methods of using ANN in safety-critical applications with the goal of creating a catalog of best practices for engineering ANNs of high quality. However, multiple overview papers exist that explicitly address particular aspects of this relationship.

Taylor et al. (4) point out the need for new paradigms of software development and certification with regard to ANNs. First, the authors state the safety-critical concerns relating to verification and validation. Subsequently, different methods applying to specific stages of the ANN lifecycle are suggested.

Kurd et al. (2) present the development of a safety case to justify the use of ANNs within safety-critical applications. They attempted to establish safety criteria for ANNs that must be enforced in safety-critical contexts. Besides that, they state challenges with respect to the trade-off between performance and safety.

Schumann et al. (1) surveyed the application of ANNs in high-assurance systems in various fields in 2010. They provide an overview of issues and challenges with respect to assurance. Furthermore, different approaches to overcoming these challenges are discussed.

Burton et al. (5) outlined existing challenges regarding safety that arise when using machine learning techniques in highly automated driving. To this end, they made use of an assurance case structure by means of the Goal Structuring Notation (6), which also demonstrates the necessity of further research activities in the context of new verification techniques. Particular focus was placed on possible methods that can be used to reduce functional insufficiencies in the perception functions based on Convolutional Neural Networks. This work has recently been extended in (7).

Falcini et al. (8) provide a general overview of deep learning (DL) and its fundamental characteristics as well as common existing automotive standards such as ISO 26262. Based on this, they performed a preliminary, lightweight applicability analysis of these standards with respect to AI-influenced automotive systems. The presented analysis shows gaps, both in terms of the development process and in terms of the product characteristics. These need to be resolved in order to allow for safety certification.

Cheng et al. (9) highlight the challenges with respect to the certification of dependable neural networks. Therefore, the inapplicability of classical engineering approaches using V-models is mentioned. Based on this, they present their considered additions to classical engineering approaches towards safety certification of an ANN. Furthermore, this concept is applied in a concrete case study of a highway ANN-based motion predictor and the resulting safety properties are evaluated.

Rao et al. (10) address specific challenges concerning functional safety during the development of deep-learning approaches for self-driving cars. They mention the challenges regarding the verification of dataset completeness for training and testing, traceability of the requirements to the ANN, and the process of transfer learning.

The existing studies have been used as an input to our study presented here. Our work contributes to the body of knowledge by combining results from these earlier papers and complementing them with comprehensive findings obtained from a systematic mapping study, yielding a comprehensive description of the current state of the art in hardening ANNs for use in safety-critical applications.

# 3 RESEARCH METHOD

In this section, the systematic mapping study is presented. We present the procedure we employed and the results we obtained. The procedure is based on the work of Petersen et al. (3), according to whom a systematic mapping study serves to "give an overview of a research area through classification and counting contributions in relation to the categories of that classification".

The objective of our study is to create a catalog of best practices that can be helpful with regard to the use of ANNs in safety-critical applications. Therefore, we will highlight the ANN characteristics that cause specific challenges regarding usage in safety-critical applications. Additionally, we will identify methods that can be helpful with regard to the use of ANNs in safety-critical applications.

To that end, we defined RQs, presented in Subsection 3.1, which are aimed to help us reach the above-mentioned goal. To answer these questions, we performed a database search for literature in this research area, which is described in Subsection 3.2. To examine the documents of interest, criteria for inclusion and exclusion were defined first. Subsequent thereto, these criteria were applied to the documents from the database search. The procedure for this is described in Subsection 3.3. Hereafter, data relevant for answering our RQs was extracted from the included documents and coded. The applied procedure is documented in Subsection 3.4. Finally, we will discuss possible threats to the validity of this mapping study in Subsection 3.5.

## 3.1 RESEARCH QUESTIONS

To guide the activities in this work, we defined two research questions (RQ), which are outlined below:

*RQ1: What are the challenges of using ANNs in safety-critical applications?*

The intention underlying this RQ was to identify challenges of using ANNs that have already arisen in safety-critical applications. In order to use ANNs in these safety-critical applications, it is necessary to be aware of such challenges, which could potentially pose a risk.

*RQ2: Which methods are proposed to overcome the challenges?*

The intention underlying this RQ was to identify methods that allow for mitigating or overcoming existing challenges regarding the quality of ANNs, with the aim being the creation of a catalog of these methods for safety-critical applications. The focus of this study was on gaining as complete an overview of these methods as possible and not on analyzing each individual method in detail. In order to obtain a certain degree of confidence on completeness, we used the approach of a systematic mapping study introduced above.

## 3.2 LITERATURE SEARCH

In order to develop the search string, an iterative approach, as mentioned in (3), was implemented. First, senior experts in the domain of safety as well as data science / Big Data were consulted to obtain existing literature that considers the use of ANNs in safety-critical applications. Based on this, keywords were extracted by a group consisting of two senior experts in the safety domain, one senior expert in the data science / Big Data domain, two doctoral candidates in the safety domain, one doctoral student in the data science / Big Data domain, and one graduate student.

Based on the aforementioned RQs, we identified meaningful keywords and an initial search string. To retrieve papers, we queried the indexing database Scopus with the search string of the respective iteration. Subsequently, we assessed the precision and noise of the query result, which required reflecting on the search string in order to increase its sensitivity and specificity. On aggregate, this process was repeated 13 times to ensure a meaningful search string.

While discussing the results of different search strings, it turned out to be reasonable to not only limit the search to "NN" but also include other synonyms of (Artificial) Neural Networks, commonly used architecture types, and keywords of different Machine Learning tasks where ANNs are typically utilized.

*Table 1 Construction of the Search String*

| *Scope* | Title, Abstract, and Keyword |
|---|---|
| *Incudes* Topic ANN | GAN, NN, auto encod[…], advers[…] net, […]conv[…] net[…], recu[…] net[…], deep learn[…],[…]reinforcement learn[…], […]supervised learn[…],[…]unsupervised learn[…] |
| *and* Topic safety-critical | mission critical, high[…] assur[…], high[…] integ[…], safety, certif[…] |
| *and* Topic challenge | challeng[…], risk[…] |

In addition, we added restrictions to the search string to reduce the workload in the subsequent "Study selection" step. Therefore, we limited the results to articles and conference papers whose subject is in the computer science area and which are published in English. Because the Scopus database only contains peer-reviewed documents – which was one of our requirements – no further steps were required in this regard.

*Table 2 Restrictions Applied on the Search*

| *Database* | Scopus |
|---|---|
| *Topic Area* | Computer Science |
| *Document Type* | Journal and conference paper |
| *Quality Criteria* | Peer-reviewed |
| *Language* | English |

Querying the database in September 2018 with the final search string resulted in a list of 885 documents:

*TITLE-ABS-KEY(("GAN" OR "NN" OR "auto encod\*" OR "\*advers\* net\*" OR "neural net\*" OR "\*conv\* net\*" OR "recu\* net\*" OR "\*deep learn\*" OR "\*reinforcement learn\*" OR "\*supervised learn\*" OR "\*unsupervised learn\*") AND ("mission critical" OR "high\* assur\*" OR "high\* integ\*" OR "safety" OR "certifi\*") AND ("challeng\*" OR "problem\*" OR "risk\*")) AND (LIMIT-TO(DOCTYPE, "ar") OR LIMIT-TO(DOCTYPE, "cp")) AND (LIMIT-TO(SUBJAREA,"COMP")) AND(LIMIT-TO(LANGUAGE,"English"))*

## 3.3 STUDY SELECTION

To limit the obtained documents to those in which the ANN is on the critical path for system safety, we defined the following inclusion, respectively exclusion, criteria and applied an iterative process:

**Inclusion criteria:** Documents were to be included if an ANN was employed in a safety-critical context. This included applications in which

- erroneous output of the ANN leads directly to harm;
- the system uses information about objects detected by an ANN as direct input to control;
- an ANN-based decision support system is used and the final decision is made by the system.

We further included documents that describe methods for preparing ANNs for use in safety-critical applications or report directly on related challenges.

**Exclusion criteria:** Documents were to be excluded if the ANN was not (intended to be) used in a safety-critical context, including the following cases:

- an ANN is used for data analysis performed after operation of the system;
- an ANN is used as a support system but not as the main algorithm described in the paper;
- the system uses information about objects detected by an ANN for tracking the objects
- an ANN-based decision support system is used and the final decision is done by a human operator

To reduce effort in the data extraction phase, it was essential to limit the false positives in the study selection phase arising due to misunderstandings among the participants regarding the accepted documents. Thus, we conducted a joint prestudy to achieve a common understanding on the above criteria among all five researchers participating in the study selection phase. This was done on the first 50 of the 885 documents, with each participant determining on the basis of the defined criteria whether to include or exclude the document in question. In this step, the title of each document was examined to decide on inclusion. In the case of ambiguity, the abstract was taken into account. The decisions were compared and different opinions were discussed. Overall, this helped to resolve minor misunderstandings regarding the definitions of the criteria.

Another measure to reduce individual bias was double-checking by two of the research participants (four-eyes principle). For this purpose, we divided the 885 documents into ten groups and assigned two participants to each of them in order to obtain two independent decisions for each instance of a document in the respective group.

Furthermore, we also ensured that the pairing of the participants assigned to a group of documents was different for each group. One of the two participants was assigned the role of the leader, and we ensured that each participant was the group leader for two groups of documents.

The overall paper selection process is shown in Figure 1. In the first step, the participants were allowed to mark a document as "tentatively accept", "tentatively reject", or "questionable" (if they could not make a concrete decision).

After all papers in a group had been marked by the two participants, the leader was responsible for the final decision. If both participants had tentatively accepted or rejected a document, it was finally accepted or excluded, respectively. Otherwise, the abstract was re-examined and a consensus was sought for the final decision. In the case of uncertainty, the paper was discussed with the other participants in a larger group meeting in order to make a decision.

Following this selection process, 185 of the 885 documents remained. A second inclusion check was subsequently made based on the full text of all included documents by a group of three participants, each performing this check individually. For that purpose, we defined the following two document categories according to our inclusion criteria:

1. Documents directly describing methods for preparing ANNs for use in safety-critical applications or reporting directly on challenges
2. Documents where ANNs are used in a safety-critical context.

The documents falling into the first category were included directly. For documents belonging to the second category, an additional check was applied. For this check, the full text was used to determine whether any direct responsibility for safe system behavior was associated with the ANN, i.e., whether an erroneous output of the ANN may cause the whole system using the ANN to transition to a hazardous state in which the occurrence of an accident is no longer controllable by the system. This criterion explicitly excluded applications having a redundant architecture with a safety channel. In such applications, it can be argued that failures of the neural network are not safety-critical and thus no special means are required to achieve sufficient quality of the ANN.

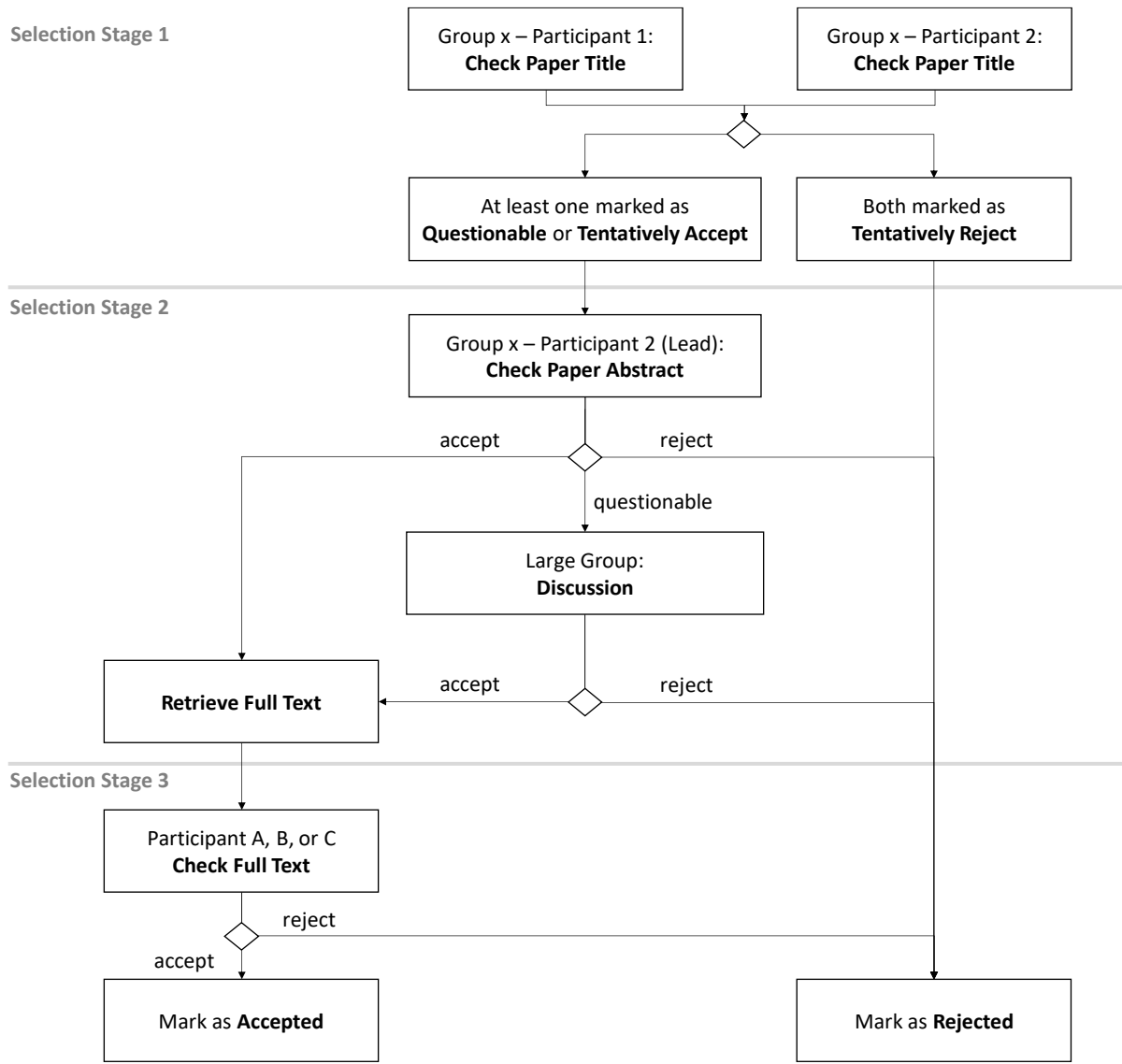

*Figure 1 Process for the Inclusion and Exclusion of Documents*

If the first condition was satisfied, it was checked whether the authors of the respective document reported at least one problem with the use of an ANN in a safety-critical application. Only if this was the case did the document pass the check and was included. This check excluded work from authors that are agnostic to the particular challenges resulting from the use of ANNs in safety-critical applications. We did not expect to be able to extract knowledge from papers of such authors.

After this second check, the number of included documents decreased from 185 to 43 documents. Moreover, during this check, there were not only exclusions due to non-compliance with the inclusion conditions – due to the unavailability of full texts, 28 documents had to be excluded in addition. Based on these 43 final documents, we conducted the data extraction and classification.

The results of each individual phase during the study selection process are represented in Figure 2.

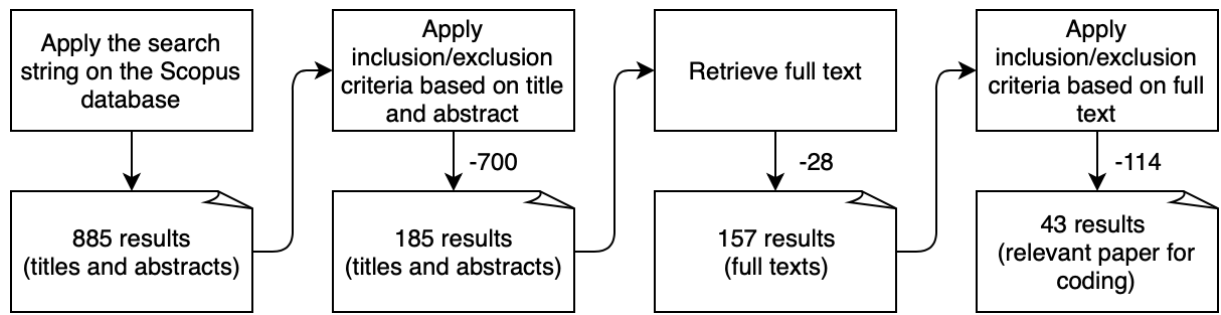

*Figure 2 Study Selection Process*

## 3.4 DATA EXTRACTION AND CLASSIFICATION

Regarding the data extraction process, the intention was to extract information from the papers that would help to answer our RQs. We used MAXQDA (https://www.maxqda.de) as a computer-assisted data analysis software that can be used for qualitative and quantitative research, to tag, respectively 'code', text passages within the documents that would help to answer our RQs. MAXQDA offers the functionality to create *codes*. These are essentially links between concepts and their corresponding paragraphs in the imported documents. The coding system of MAXQDA allows structuring and organizing information by creating hierarchical structures of codes. These can again consist of sub-codes, which may link to one or more paragraphs. Additionally, MAXQDA enables the creation of 'memos' as part of the codes and documents. These can be used to provide supplementary information in textual form to the corresponding element.

In order to create an appropriate hierarchy for answering the RQs, each participant created and linked codes to paragraphs directly when examining a document. Regular meetings were required to compare the created codes mutually and to synchronize them. In addition, based on these low-level codes, clusters were created, which were abstracted with codes on a higher hierarchy level. An alternative approach – working with code templates – was rejected due to the great effort that would have been necessary to perform it.

For the data extraction process, the documents were distributed among four participants. To allow parallel extraction, a separate MAXQDA file was created for each participant. Each participant imported only the assigned documents. In order to obtain a basic uniform structure for the codes and to enable easy consolidation later, we pre-defined different types of codes.

Initially, the following two types of basic codes were defined based on the RQs: 'Challenge' and 'Method'. These were extended with the types 'Characteristics', 'Application', and 'Definition'. The basic codes were defined to address the RQs directly. The additional code type 'Application' was intended to obtain insights into the application areas in which ANNs are used. The code type 'Characteristics' was intended to identify the specific properties of ANNs that motivate the named challenges (e.g., non-convexity, non-linearity, non-deterministic results of training), and the code type 'Definition' was used to create definitions of common terminology as well as common abbreviations in the context of ANNs.

We used the following standardized coding notation for codes: *[CodeType]_[CodeName]*

In this, *[CodeType]* refers to one of the five defined types and *[CodeName]* provides, if available, either a term given by the author or a concise and descriptive name.

Moreover, we used memos to document the relationship between characteristics and challenges, as well as between challenges and methods. To this end, we named the code of a characteristic or the code of a challenge in the memo of a challenge code or a method code, respectively. To ensure that each participant used concise and descriptive names for their codes, weekly meetings were held. During these, the created codes as well as the memos were reviewed mutually under these aspects.

Since 'Characteristics' were rarely mentioned in the reviewed documents and even more seldom related to specific 'Challenges', we decided to exclude them from further analysis.

The next step was to find suitable clusters for all code types except the definition codes. These clusters should group as many similar codes as possible and the clusters should differ significantly from each other. These clusters are the particular challenges and methods that will be present below. For this purpose, each participant first thought about which cluster would be best suited for the codes they had created themselves. Separate meetings for the different code types were held among the participants to brainstorm the individual results. We proceeded in such a way that the participant who had most of the clusters for the corresponding code type presented them to the team, explaining which codes fall into the respective cluster and to what extent the clusters differ from each other. In the same way, changes were then proposed for these by the other participants, if considered necessary. Afterwards, the participants put their codes into the suggested clusters. If there were outliers that did not seem to fit any existing cluster, either a missing cluster was suggested directly by the respective participant or these codes were included in a separate list for which an attempt was made to find suitable clusters.

Once this process had been completed, the clusters formed were given summary descriptions and then presented to senior experts, who assessed them in terms of comprehensibility and plausibility. Appropriate proposals for adjustments were then made where required.

## 3.5 VALIDITY EVALUATION

This subsection summarizes the threats to the validity of the presented study we consider most crucial. Their discussion in the following is structured according to the steps of the research method we applied.

**Research Questions:** In general, RQs may not be entirely helpful for realizing the goal of a study. To address this threat to validity, they were assessed by senior experts from the domain of safety as well as data science / Big Data. As a result, the initial RQs were corrected once, taking the noted weaknesses into consideration. The final evaluation by the same experts did not require any further revision.

**Literature Search:** The creation of the search string was based on keywords we extracted from literature preselected by senior experts. Therefore, the creation of the string might be prone to bias. For this reason, we conducted an iterative procedure with a quality assessment. The quality assessment with regard to precision and noise was performed by means of random sampling within the documents of the respective database search results. In the samples, documents that could negatively influence precision and noise may have been underrepresented.

Moreover, contrary to the advice by Petersen et al. (3), we limited your search for relevant documents to the Scopus database, ignoring IEEE, ACM, and Inspec. This may have resulted in the omission of some relevant documents. However, we decided to limit our search to the Scopus database as past studies revealed a great overlap between the different databases.

**Study Selection:** As the decisions to include or exclude a document may be biased by the individual participants, we tried to achieve a common understanding regarding the inclusion and exclusion criteria among all five participants during the study selection phase. This was done by means of a pre-study conducted on a subset of 50 documents selected from the retrieved 885 titles. The results of the pre-study indicated the need for clarification with regard to the definition of the inclusion and exclusion criteria. This was achieved by discussing example documents for both cases, inclusion and exclusion. Furthermore, decisions regarding the selection were made by groups of two for the first selection stage.

A further threat is the circumstance that for 28 documents, the full text could not be obtained, which means that approximately 15% of the documents classified as potentially relevant based on their title and abstract could not be further inspected.

**Data Extraction and Classification:** Throughout this process, the data was extracted initially by creating and linking codes within each document in MAXQDA. To this end, the documents to be processed were distributed among the participants without interleaving. Due to the high number of papers that had to be coded, there was no second inspection of the extracted data by any other participant. This may have resulted either in important paragraphs being omitted or in incorrect linking between codes and paragraphs. Only the comprehensibility and the plausibility of the generated codes were checked in regular meetings, including mutual inspection of those codes.

Starting from these codes, clusters were established. In order to find clusters, the participants suggested possible candidates based on their individual codes. It is possible that wrong candidates were suggested based on some bias of the participants. In addition, the assignment to the clusters may have been wrong due to incorrect interpretation of its scope. Consequently, the relationship between clusters of different code types, which is based on the respective code memos in the clusters, might be incorrect. Since any inaccuracies during this process have a direct impact on the later analysis and thus on the results, we did perform countermeasures. During the clustering process, each participant presented the meaning of the codes intended for the clustering. This enabled the other participants to propose corrections if considered necessary. Furthermore, senior experts assessed the plausibility of the identified relationships between methods and challenges as well as between challenges and characteristics.

## 4  RESULTS

Based on our classification, we identified the application domains in which ANNs are currently being discussed for safety-critical applications. For each paper, we collected the application domain discussed, provided it was mentioned in the paper. In total, 35 papers explicitly named an application domain, with four domains – *autonomous driving*, *aviation*, *industrial robots*, and *medicine* – being mentioned more than once (Figure 3). In this context, it became apparent that ANNs are discussed comparatively frequently in the safety-critical areas of *autonomous driving* and *aviation*. Application domains mentioned by only one paper are mining, railway, energy system, bridge crack detection, malware detection, automotive engine calibration, use of semantic segmentation, and industrial hoist mechanism.

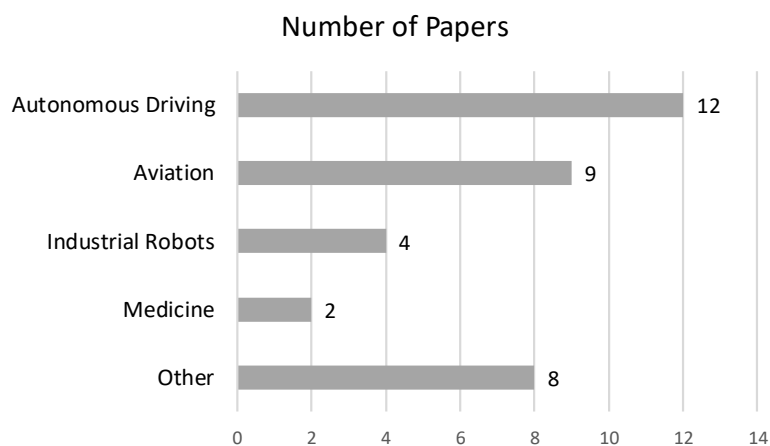

*Figure 3 Application Domains Mentioned in Mapping Study*

As an answer to RQ1 (*What are the challenges of using Artificial Neural Networks in safety-critical applications?*), we extracted the challenges represented in Table 3 from the final set of 43 papers. The table contains a short name and a brief description for each challenge. We derived these names and descriptions from the terms used regularly in the relevant papers. Because of our systematic approach to extracting the data in this study, we have confidence in the completeness of Table 3 as a list of the challenges that are currently within the scope of researchers developing ANNs for safety-critical applications. Consequently, we need a set of engineering methods that cover at least all of these challenges and provide sufficient solutions if a network is supposed to be used in a safety-critical application.

*Table 3 Challenges Identified in Mapping Study*

| Challenge | Description |
|---|---|
| Interpretability of ANN | Understand the decision taken by the ANN |
| ANN Development Paradigm | Deal with the development process of NN-infused components |
| Appropriately Distributed Training Data | Create a training dataset that is appropriately distributed over the anticipated operation domain |
| Noisy Real World Data | Deal with noise during operation and with noisy training data |
| Generalization of ANN | Handle challenges like overfitting of the ANN |
| Convergence of ANN | Deal with local minima (non-convexity) and ensure convergence of an ANN during the training process |
| Assurance of Safety-related Properties | Assure safety-related properties of the ANN |
| Runtime Safety | Evaluate and mitigate risk during the operation of the system originating from the ANN |
| Verification and Validation of ANN | Use appropriate verification and validation (V&V) methods for ANN |
| Robustness against Adversarial Attacks | Deal with intentional manipulation of input data during operation to trigger an ANN malfunction |

As an answer to RQ2 (*Which methods exist to overcome the challenges?*), we found the method clusters presented in Table 4. In this table, we list all the extracted methods that address at least one challenge cluster. We distinguish between mappings based on the literature and mappings performed by a project team. The former only comprise challenges explicitly mentioned in the included documents. The latter comprise challenges that the participants of the study perceived as being addressed by this method during extraction. The mappings by project teams merely extend the mappings from the literature; the reason for including them was to ensure that each method is mapped to at least one challenge.

The goal of Table 4 is to give a complete overview of the different methods for increasing the quality of neural networks that we identified with the conducted mapping study. The rows in Table 4 group work that, according to our understanding, is very similar. It is, however, not our goal to introduce generally accepted names or clusters for existing methods, and the referenced literature could also be grouped differently. The description field can only give a very brief explanation of the method and the reader should consult the referenced papers for more information.

*Table 4 Methods Identified in Mapping Study*

| Method | Description | Challenges (Literature) | Challenges (Project Team) | References |
|---|---|---|---|---|
| **Adaptive Resonance Theory** | Usage of Adaptive Resonance Theory to solve the plasticity vs. stability dilemma of ANNs. | Generalization of ANNs | | (1) |
| **Advanced Data Collection Strategy** | Application of strategies such as simulation and automated data generation to collect data for training and thereby address data representativeness and uniformity challenges. | Appropriately Distributed Training Data | | (9), (11) |
| **Adversarial Deep Learning** | Application of adversarial deep learning as training data augmentation by adding existing images that are minimally disturbed to increase robustness against adversarial attacks. | Generalization of ANNs, Assurance of Safety-related Properties | Robustness against Adversarial Attacks | (12), (13), (14) |

| Method | Description | Challenges (Literature) | Challenges (Project Team) | References |
|---|---|---|---|---|
| **Assurance Case for ANNs** | Creation of an assurance case approach by decomposing the safety goals of the system into technical performance requirements in the Machine Learning function with explicit consideration of assumptions on the system and its environment. | Verification and Validation of ANNs, Assurance of Safety-related Properties | | (5), (2) |
| **Automatic Optimizing of ANN Configuration** | Use of automatic optimization of ANNs: automatically changing the number of neurons in hidden layers; selection of best architecture based on comparison to help in ANN development and achieve better generalization. | | ANN Development Paradigm, Generalization of ANNs | (15), (4) |
| **Bayesian Deep Learning** | Obtaining the uncertainty of an ANN's decision as output from the ANN and using a Bayesian method for modeling uncertainty to improve confidence in the output. | Verification and Validation of ANNs | | (16) |
| **BPTT Algorithm** | Use of the BPTT algorithm to learn from training data that is affected by noise. | Noisy Real-World Data | | (17) |
| **Computing Maximal Bounds for the Resilience of ANNs** | Establishment of maximum and verified bounds for the resilience of given ANNs against input disturbances in order to investigate how much noisy or even maliciously manipulated sensory input is tolerated. | Noisy Real-World Data, Robustness against Adversarial Attacks | | (18) |
| **Conversion to Decision Tree** | Conversion of ANN into a decision tree to provide additional insight into the inner workings of the network. | Verification and Validation of ANNs | Interpretability of ANNs | (19), (4) |
| **Cross-Validation** | Use of cross-validation for improved generalization during training. | | Generalization of ANNs | (20), (21), (14) |
| **Data Augmentation** | Use of data augmentation to help the model learn image-invariant features through geometric transformation, hence increasing variance and thereby moving towards uniformity of the dataset. | Appropriately Distributed Training Data | | (14) |
| **Data Preprocessing** | Pre-processing of the input data to enhance training. | Generalization of ANNs | | (22) |
| **Data-Type with Adequate Dynamic Value Range** | Use of a data type providing a just-enough dynamic value range and precision to support the design of a network that facilitates dealing with noisy data. | Noisy Real-World Data | | (23) |
| **Dedicated ANN Development Process** | Use of a dedicated ANN development process that incorporates V&V in the development lifecycle. | | ANN Development Paradigm | (24) |
| **DeepXplore** | Use of DeepXplore as an efficient whitebox testing framework for large-scale DL systems. | Verification and Validation of ANNs | | (13) |
| **Differential Testing for ANN** | Use of multiple DL systems with similar functionality as cross-referencing oracles to identify | Verification and Validation of ANNs | | (13) |

| Method | Description | Challenges (Literature) | Challenges (Project Team) | References |
|---|---|---|---|---|
| | erroneous edge cases without manual checks. | | | |
| **Distraction Reporting** | Training of a neural network to respond correctly to events involving distractions and to report, as status output, the detection of ignored distractions. This can help to deal with noisy data. | Noisy Real-World Data | | (19) |
| **Dynamic Monitoring of ANN** | Use of tools that can dynamically monitor the quality of the neural network and its internal parameters. | Runtime Safety | | (1) |
| **Energy Function Monitoring** | Use of the energy function of an error as an indicator of the performance of the ANN. | Runtime Safety | | (25) |
| **Equivalence-Class-based Black-Box Testing** | Use of equivalence-class-based testing techniques for targeted testing of the software component containing the learned function, including the use of systematically selected test data. | Verification and Validation of ANNs | | (5) |
| **Performance Evaluation** | Use of methods such as F-score, precision, recall, or special performance indices to evaluate ANN performance. | | Verification and Validation of ANNs | (14), (11), (1) |
| **Exception Training** | Use of training by exposing a model to situations beyond what is considered likely or even realizable in a real-life setting; a form of stress training and testing that will improve the model's capability to ignore noise. | Noisy Real-World Data | | (19) |
| **Formal Analysis of an ANN** | Application of formal methods such as static analysis or symbolic reasoning in order to guarantee properties. | Assurance of Safety-related Properties, Verification and Validation of ANNs | | (9) |
| **Global Optima Consideration** | Search for global optima instead of local optima during the learning process, with adjustment of the weights of the neural network in order to yield better results. | Convergence of ANNs, Verification and Validation of ANNs | | (26), (27), (28) |
| **Handling of Rules in Online Learning ANNs** | Application of rules/regulations by monitoring the learning and functioning of the OLNN, or representing them within the OLNN, making OLNN self-regulating. | Runtime Safety | | (19) |
| **Handling of Soft Errors** | Identification of vulnerable static instructions and selective duplication of the instructions to protect from soft errors. | | Assurance of Safety-related Properties, Runtime Safety | (23) |
| **Lyapunov Stability Monitoring** | Implementation of Lyapunov stability-based monitors to capture unstable behavior as a dynamic monitoring tool. | Runtime Safety | | (1), (4) |
| **Model Parameter Visualization** | Visualization of model parameters to gain insights into the ANN and the operation of the classifier. | Interpretability of ANNs | | (16) |

| Method | Description | Challenges (Literature) | Challenges (Project Team) | References |
|---|---|---|---|---|
| **Natural Language Explanation of Learned Features** | Generation of a natural language explanation referring, in human-understandable terms, to the contents of the input image in order to explain which features were relevant for the classification. | Interpretability of ANNs | | (5) |
| **Neural Network Description Languages** | Use of neural network specification languages that allow the abstract specification of neural networks and a corresponding compiler that translates the specification into C++ network classes for better documentation. | | Verification and Validation of ANNs | (4) |
| **Neuron Coverage Criteria** | Application of neuron coverage, which at a high level is similar to the code coverage of traditional systems, to argue on the completeness of the testing process of ANNs. | Verification and Validation of ANNs | | (13) |
| **Normalization Layer** | Addition of a normalization layer to reduce the impact of faults by averaging the faulty values with adjacent correct values, hence improving generalization. | Noisy Real-World Data | Generalization of ANNs | (23) |
| **ANN Performance Verification** | Verification of the performance of an ANN. | Verification and Validation of ANNs | | (24) |
| **Picasso** | Use of Picasso as a tool that supports the development process of neural networks and helps to monitor and understand the learning process of neural networks. | ANN Development Paradigm | | (10) |
| **Piecewise Linearization of Activation Function** | Application of piecewise linearization of nonlinear activation functions to facilitate the verification of neural networks. The nonlinearity of the activation functions is one source of complexity when verifying neural networks. | Verification and Validation of ANNs | | (26) |
| **Relevance Visualization** | Generation of a relevance visualization in order to highlight portions of an image that strongly influence classification results. | Interpretability of ANNs, Appropriately Distributed Training Data | | (5), (16), (29) |
| **Resilient Design** | Implementation of a resilient design to make the system robust against disturbances. | Runtime Safety, Noisy Real-World Data | | (30), (4), (31), (1), (32) |
| **Risk-Aware Resampling of Training Data** | Use of a risk-aware resampling technique in order to combat the data mismatch problem and heterogeneous costs. | Appropriately Distributed Training Data | | (33) |
| **Rule Extraction** | Extraction of rules that can be used for validation against requirements. | Verification and Validation of ANNs, ANN Development Paradigm, Interpretability of ANNs, Assurance | | (19), (9), (24), (4), (1) |

| Method | Description | Challenges (Literature) | Challenges (Project Team) | References |
|---|---|---|---|---|
| | | of Safety-related Properties | | |
| **Rule Initialization** | Initialization of the network with a starting point from which the network can adapt, thus improving confidence in its behavior. | Runtime Safety | | (4) |
| **Rule Insertion** | Periodic use of rule insertion, while in operation or offline, to steer a dynamic network towards a desired operational regime. | Runtime Safety, ANN Development Paradigm | | (4), (34), (10) |
| **Safe Controller Learning** | Use of an inherently safe parametric controller as a basis for the learning task in order to restrict learning to permitted behavior. | Appropriately Distributed Training Data | | (35) |
| **Safety Bag** | Use of an additional source of evidence to minimize the impact of functional insufficiencies of an ANN by means of runtime measures that make use of secondary channels not used by the Machine Learning function. | Runtime Safety, Assurance of Safety-related Properties, Appropriately Distributed Training Data | Verification and Validation of ANNs | (5), (35), (4), (36), (32) |
| **Safety Margin of Operational Scope** | Use of safety margins to limit the outcome of system operations to a certain scope. This ensures that operation outcomes will not fall within critical regions, which might result in system failure. | Verification and Validation of ANNs, Appropriately Distributed Training Data | | (26), (30), (37) |
| **Self-Correcting Iteration Times during Learning** | Use of self-correcting iteration times during learning in order to extend the iteration times depending on the current training and testing error. This contributes to avoiding deficiencies in learning. | Generalization of ANNs | | (28) |
| **Separation of Data** | Separation of the available data into three sets: training, validation, and testing sets. | Verification and Validation of ANNs | | (13), (4), (1) |
| **Stable-ANN** | Use of stable-ANNs, which are bounded in a finite interval and have proven stability and convergence. | | Convergence of ANNs | (38), (39) |
| **Static ANN Analysis** | Application of static analysis techniques to neural networks to provide a mapping between several independent variables and a dependent variable. | | Verification and Validation of ANNs | (4) |
| **Transfer Learning** | Use of transfer learning (training a neural network on one task and using part of the pre-trained weights for another related task) to enhance ANN development. | ANN Development Paradigm | | (16), (10) |
| **Uncertainty Quantification** | Consideration of uncertainty within the model in order to check whether assumptions on the model or the model itself might not represent the reality accurately enough. | Verification and Validation of ANNs, Appropriately Distributed Training Data, Generalization of ANNs, Assurance | | (5) |

| Method | Description | Challenges (Literature) | Challenges (Project Team) | References |
|---|---|---|---|---|
| | | of Safety-related Properties | | |
| **Verification for Pre-Trained Sigmoidal Nets** | Use of the verification method of pre-trained sigmoid nets as a verification method that is accepted by US aviation authorities as a valid basis for certification of software containing pre-trained, static ANN modules. | Verification and Validation of ANNs | | (40) |
| **Verification Frameworks for ANN** | Utilization of neural network verification frameworks that abstract linear arithmetic constraints to Boolean combinations. | Assurance of Safety-related Properties, Verification and Validation of ANNs | | (20), (12) |
| **Verification of ANN with Monte-Carlo Estimation** | Use of Monte Carlo estimation to estimate the probability of failures for the verification of a neural network. | Verification and Validation of ANNs | | (40) |
| **W-Model Development Paradigm** | Application of a W-model to conceptually integrate a V-model for data development into the standard V perspective of software development. | Verification and Validation of ANNs, Assurance of Safety-related Properties | | (8) |

## 5 DISCUSSION

The usage of ANNs in safety-critical applications requires assuring that their probability of producing faults is sufficiently low. Reasons for producing faults are the challenges identified in this study. Methods that assist in keeping the network from failing are the engineering methods identified in this study. Consequently, we need to assure that all challenges mentioned in Table 3 are sufficiently addressed by the engineering methods in Table 4. In other words, the hardening of the network has to be done in such a way that all challenges are resolved. To give an overview of the status of the mapping between challenges and methods, we list the numbers of methods for each challenge in Table 5.

*Table 5 Number of References to Challenges in Mapping Study*

| Challenge | # Mappings in Literature | # Mappings in Literature + Project Team |
|---|---|---|
| Interpretability of ANNs | 4 | 5 |
| ANN Development Paradigm | 4 | 6 |
| Appropriately Distributed Training Data | 8 | 8 |
| Noisy Real-World Data | 7 | 7 |
| Generalization of ANNs | 5 | 8 |
| Convergence of ANNs | 1 | 2 |
| Assurance of Safety-related Properties | 8 | 9 |
| Runtime Safety | 8 | 9 |
| Verification and Validation of ANNs | 19 | 23 |
| Robustness against Adversarial Attacks | 1 | 2 |

We can see in Table 5 that all challenges are addressed by methods. A lot of work has been invested in the field of verification and validation of ANNs, while security-related robustness against adversarial attacks has not been investigated too much in the context of safety-critical systems until now.

We can thus observe that methods do exist that allow addressing the specific challenges of using ANNs in safety-critical applications. It is an additional question, beyond the scope of this work, whether these methods address the challenges sufficiently well and how a process for the systematic engineering and hardening of ANNs can be designed based on the available methods.

# 6  CONCLUSION

In 1968, the rising importance of software in technical systems and the lack of experience and commonly accepted methods in the development of this software led to a new discipline of computer science: Software Engineering. Since then, great steps have been taken and whole research institutes have evolved around this topic. In the last decade, the relevance of Machine Learning in general and ANNs in particular for software-intensive systems has grown dramatically. Just as with traditional software, we need to assist the development of ANNs with proven engineering methods and adequate processes in order to create networks of sufficient quality for use in safety-critical applications. The way ANNs are developed and how they behave differs significantly from what we learned about software engineering in the last fifty years. Therefore, we have to rethink what proven software engineering principles mean for the development of ANNs, what processes need to be defined, and which methods are most appropriate to develop them in a way that satisfies our quality expectations. With this work, we have contributed initial input, derived systematically from existing articles, for the development of such methods and processes and for the extension of the scope of Software Engineering through the development of ANNs. For now, the results of this study are intended to create awareness for the variety of existing methods and can assist in choosing methods that increase the quality of networks to be used in critical applications.

**Acknowledgments.** Parts of this work have been funded by Ministry of Science, Education, and Culture of Rhineland-Palatinate in the context of MInD project and the German Federal Ministry of Education and Research (BMBF) under grant number 01IS16043E (CrESt).